\documentclass[final,5p,times,twocolumn]{elsarticle}

\usepackage{lineno,hyperref}
\modulolinenumbers[5]

\usepackage{graphicx}
\usepackage{xspace}
\usepackage{amsmath}
\usepackage{amssymb}
\graphicspath{{./plots/}}

\def\xmax{\ensuremath{X_{\rm max}}\xspace}

\def\nmu{\ensuremath{N_\mu}\xspace}
\def\mnmu{\langle \ensuremath{N_\mu}\rangle\xspace}

\newcommand{\sibyll}{{\sc Sibyll~2.3c}\xspace}

\def\qgs{\textsc{QGSjet}\,II-04\xspace}
\def\epos{\textsc{Epos-lhc}\,\xspace}
\def\fluka{\textsc{Fluka}\,\xspace}

\def\corsika{\textsc{Corsika}\,\xspace}
\def\conex{\textsc{Conex}\,\xspace}

\newcommand{\figsize}{0.90}
\newcommand{\figsizeL}{0.95}

\begin{document}

\begin{frontmatter}

\title{Probing the energy spectrum of hadrons in proton air interactions at ultrahigh energies through the fluctuations of the muon content of extensive air showers}

\author{Lorenzo Cazon}
\address{Laborat\'{o}rio de Instrumenta\c{c}\~{a}o e F\'{i}sica Experimental de Part\'{i}culas (LIP) - Lisbon, Av.\ Prof.\ Gama Pinto 2, 1649-003 Lisbon, Portugal}

\author{Ruben Concei\c{c}\~{a}o}
\address{Laborat\'{o}rio de Instrumenta\c{c}\~{a}o e F\'{i}sica Experimental de Part\'{i}culas (LIP) - Lisbon, Av.\ Prof.\ Gama Pinto 2, 1649-003 Lisbon, Portugal}
\address{Instituto Superior T\'ecnico (IST), Universidade de Lisboa, Av.\ Rovisco Pais 1, 1049-001 Lisbon, Portugal}

\author{Felix Riehn}
\ead{friehn@lip.pt}
\address{Laborat\'{o}rio de Instrumenta\c{c}\~{a}o e F\'{i}sica Experimental de Part\'{i}culas (LIP) - Lisbon, Av.\ Prof.\ Gama Pinto 2, 1649-003 Lisbon, Portugal}

\date{\today}
\begin{abstract}
  We demonstrate that the shower-to-shower fluctuations of the muon content of extensive air showers correlate with the fluctuations of a variable of the first interaction of Ultra High Energy Cosmic Rays, which is computed from the fraction of energy carried by the hadrons 
  that sustain the hadronic cascade. The influence of subsequent stages of the shower development is found to play a sub-dominant role. 
As a consequence,  the shower-to-shower distribution of the muon content is a direct probe of the  hadron energy spectrum of interactions beyond those reachable in human-made accelerators.
\end{abstract}


\end{frontmatter}


\section{\label{sec:intro}Introduction}

Ultra-High-Energy Cosmic Rays (UHECRs) collide with nuclei of the Earth's atmosphere at center-of-mass energies that surpass the LHC energy scale, creating huge cascades of particles, the Extensive Air Showers (EAS)~\cite{EngelReview}. EAS constitute an unique experimental opportunity to explore hadronic interactions well beyond the energy scale attained by the largest human-made accelerator, the Large Hadron Collider (LHC). 

On the other hand, the precise composition of UHECR is still unknown, as the nature of the primaries has to be inferred from EAS measurements themselves by using phenomenological models of the hadronic interactions tuned to a limited range in energy and kinematic phase-space.
Breaking this degeneracy is one of the challenges currently faced by UHECR physics. This is to be achieved by increasing the number of independent observables taken from EAS.

Muons are direct messengers from the hadronic activity within EAS. The overall muon number can signal inconsistencies in the shower description. In fact, measurements at the Pierre Auger Observatory have shown that simulations, using the LHC-tuned hadronic interaction models, underestimate the muon content in air showers~\cite{HAS,PRLhad}. It has also been shown that the shower-to-shower variation in the muon content is useful information in discerning the composition of UHECR primary~\cite{UngerKH}. As a consequence, the effort to better measure EAS muons is increasing and several experiments are deploying upgrades to their detectors to enhance the sensitivity to them (see for instance~\cite{Aab:2016vlz}). 

It is thus necessary to deepen research into the phenomenology of the muon component and in particular to understand the shower-to-shower distributions.

In this letter, we demonstrate that the fluctuations of the muon content of EAS directly correlate with the fluctuations of a variable that probes the hadron energy spectrum of the first interaction. The subsequent shower interactions give the overall scale of the muon content, but play a sub-dominant role in the shower-to-shower fluctuations.

\section{\label{sec:nmu-dis} The shower-to-shower fluctuations in the muon content of proton-induced EAS}

\begin{figure}
    \centering
  \includegraphics[width=\figsize\columnwidth]{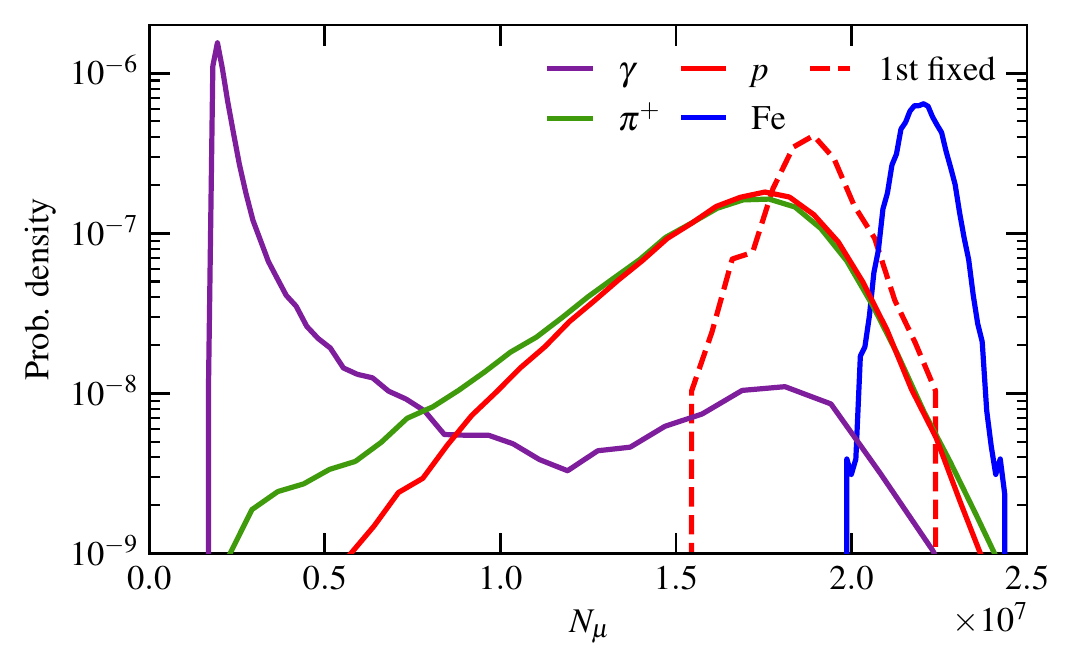}
  \caption{\label{fig:nmu-dis}Distribution of the total number of muons produced in simulated EAS initiated by different primaries. The results for a proton primary, where the multiparticle production of the first interaction is fixed, is shown by the dashed line. Simulations were done with \conex using \epos and \fluka as high and low energy interaction models respectively.}
\end{figure}

EAS initiated by UHECRs develop in two components: the hadronic and the electromagnetic (EM) cascades. The EM component consists of photons, electrons and positrons. It arises from the EM decay of mesons in the hadronic component, predominantly from neutral pions, $\pi^0$. Photons and electrons in the EM cascade only rarely undergo interactions that produce hadrons. In each interaction, approximately a third of the energy is transferred from the hadronic to the EM cascade, corresponding to the $\pi^0$ production rate. As the energy within the hadronic cascade decreases, it eventually becomes more likely that unstable hadrons decay rather than interact, this being the stage of the hadron cascade where most muons are formed~\cite{EngelReview,Meurer:2005dt}.

In Fig.~\ref{fig:nmu-dis} the shower-to-shower distributions of the number of muons, $\nmu$, for ensembles of simulated air showers initiated by different primaries are shown. The primary energy is $10^{19}\,$eV and the zenith angle is $67^\circ$. 
We define $\nmu$ as the number of muons with $E>1\,$GeV arriving at ground.
If one considers lower zenith angles the truncation of the muon production distribution and the muon propagation~\cite{TransportModel} change the resulting distributions, although their main characteristics remain the same (see~\ref{app:zenith}).

At $10^{19}\,$eV, air showers typically contain $10^7$ muons with $E>1\,$GeV. In Fig.~\ref{fig:nmu-dis}, one sees that the distribution differs substantially between different primaries. The relative fluctuations $\sigma[\nmu]/\left<\nmu\right>$ for proton primaries are $\sim 0.15$ (post-LHC model average\footnote{The variation across interaction models of the relative fluctuations for proton primaries is $8\,\%$. See~\ref{app:models} for details.}. In general, this value is much larger than $1/\sqrt{\langle \nmu \rangle}=10^{-4}$, which is what one would obtain if the muons in the EAS were produced independently. Instead, the relative fluctuations indicate that the number of independent participants is $1/0.15^2 \sim 44$, a value of the same order as the number of particles typically produced in a single hadronic interaction. This suggests that the fluctuations could be dominated by phenomena in the first interaction. 
This idea has been tested with a modified air shower simulation wherein the first interaction has been fixed and showers were simulated by reinjecting all secondaries from this particular first interaction. The resulting effect on the distribution of the number of muons is shown in Fig.~\ref{fig:nmu-dis} by the dashed line. It can be seen that the width of the distribution is reduced while the average remains the same. The relative fluctuations in this modified simulation are much lower, around $0.05$.

A simple yet effective picture to understand this in terms of the shower dynamics is provided by Heitler-Matthews (HM) type models~\cite{HM}. In such models, air showers are described as codeveloping EM and hadronic cascades, where all hadronic interactions yield a constant multiplicity. The produced particles are separated into two groups, those that decay into EM particles ($\pi^0$), feeding the EM cascade, and those that remain in the hadronic cascade ($\pi^{\pm}$), accounting for the hadronic multiplicity. In each interaction, energy is equally shared by all produced particles. The variable describing the development of the shower is the generation number, $i$, which counts the number of interactions a particle has undergone. Particles arising from the first interaction are generation $i=1$. Due to the equipartition of energy and to the transfer to the EM cascade, the energy in the hadronic cascade rapidly decreases. When it falls below a critical energy, $E_c$, the cascade is terminated and the hadronic particles decay into muons. The average number of muons in the shower can be written as a function of the total ($m_{\rm tot}$) and hadronic ($m$) multiplicity:
\begin{equation}
  \langle N_{\mu}(E) \rangle ~=~ m^g ~=~ \mathcal{C} \, E^{\beta} \ .
  \label{eq:avg-nmu}
\end{equation}
Here $g$ is the \textit{critical generation} number, $E$ is the energy of the primary particle, $\mathcal{C}=E_c^{-\beta}$ is a normalization constant,  and $\beta=\ln{m}/\ln m_{\rm tot}$.
We have set $\beta$ to $0.93$, taken from post-LHC hadronic interaction models (\ref{app:models}).

In order to understand the shower-to-shower fluctuations of $N_\mu$, one must consider that the multiplicity varies from interaction to interaction.
The average hadronic multiplicity in generation $i$ can be calculated as $m_i~\equiv~N_i/N_{i-1}$, where $N_i$ ($N_{i-1}$) is the number of interacting hadrons in generation $i$ ($i-1$).
The number of muons in the shower is then $N_{\mu}=\prod_{i=1}^g{m}_i$. Note that, while the critical generation, $g$, is well defined for the HM model, allowing fluctuations of the hadronic multiplicity changes the overall energy budget of each sub-shower, leading to the fluctuation of the generation where the energy threshold $E_c$ is reached. Therefore, $g$ must now be understood as an average parameter for the whole shower.

Assuming the hadronic multiplicities of all interactions arise from a common probability distribution with mean, $m$, and dispersion, $\sigma(m)$, the fluctuations of $m_i$ are given by
\begin{equation}
  \sigma( {m}_i )~=~\frac{\sigma(m)}{\sqrt{N_{i-1}}} \ .
  \label{eq:fluct-gen}
\end{equation}
We thus find that the fluctuations of the average multiplicity in generation $i$ are suppressed by the number of interactions resulting from the previous generation $N_{i-1}$. As the number of particles/interactions grows exponentially with the number of generations, the contributions to the fluctuations from later shower stages become increasingly smaller. This behaviour is typical for cascade processes with fixed multiplicity, as they occur in photomultiplier tubes~\cite{prescott1966,tan1982}.

In addition, for realistic multiplicity distributions, $\sigma(m)$ decreases with energy~\cite{Alner:1984is}. As the interaction energy decreases from one generation to the next, fluctuations from later stages are further suppressed. We conclude that the overall fluctuations in the number of muons are dominated by the fluctuations in the first interaction, and that the contributions from further generations are exponentially suppressed.

So far, fluctuations of the shower were explained by fluctuations of the particle multiplicity. Nevertheless, an additional source of fluctuations comes from the fact that in each interaction, energy is shared among the emerging particle in a uneven and stochastic way. 
Taking  into account only those fluctuations arising from the first interaction, one can write the total number of muons as the sum of the average number of muons that are produced in the $m_1$ subshowers that come out of the first interaction, i.e.\
\begin{equation}
  N_{\mu,1}(E)~=~\sum_{j=1}^{m_1} \langle N_{\mu}(E_j) \rangle = \sum_{j=1}^{m_1} \mathcal{C} \, E_j^\beta \ ,
  \label{eq:master}
\end{equation}
where the subscript $1$ in   $N_{\mu,1}$ denotes that only the fluctuations in the first interaction have been accounted for, and $E_{j}$ denotes the energy carried by the $j$th particle.  We have used the fact that nucleon and pion initiated showers produce similar number of muons, as shown in  Fig.~\ref{fig:nmu-dis}.
We define $x_{j}=E_{j}/E$ as the fraction of the primary energy carried by particle $j$. Each sub-shower is thus weighted by $x_j^{\beta}$ in the final number of muons.
Defining
\begin {equation}
  \alpha_1 \equiv \sum_{j=1}^{m_1} x^\beta_{j} \ ,
\end{equation}
  we find that the number of muons in a shower is related to the average number of muons
  $N_{\mu,1}(E)~=~ \alpha_1 \, \langle N_{\mu}(E) \rangle \ .$
For $\beta=1$, the variable $\alpha_1$ represents the fraction of energy that is passed on to the hadronic cascade. Its distribution is given by the hadronic energy spectrum~\cite{Sirunyan:2017nsj}. In the opposite case ($\beta=0$), $\alpha_1$ becomes $m_1$, so $\beta$ shifts the weight between energy and multiplicity. In contrast to the HM model, both the multiplicity and energy fluctuations are included.

  We can also introduce the fluctuations induced by the second generation, writing $N_{\mu,2}(E)=\sum_{j=1}^{m_1}\sum_{k=1}^{m_{2j}} \mathcal{C} \, E_{jk}^\beta $ where $E_{jk}=x_jx_{jk}E$ is the energy carried by second generation particles, where $j$ and $k$ run through the different combinations of 1st and 2nd generation particles respectively.
This procedure can be generalized to account for any number of generations. Given $\alpha_1$ we can define $\alpha_i$ recursively for any generation as $\alpha_i \equiv N_{\mu,i} / N_{\mu,i-1}$ and translate the sum of sums into a product. Including fluctuations up to generation $g$ the number of muons is given by $N_\mu = \langle N_\mu(E) \rangle \, \prod_{i=1}^g \alpha_i$. As before, the number of particles in the hadronic cascade increases with generation so the fluctuations in $\alpha_i$ decrease. By grouping fluctuations of all the generations but the first in a single parameter $\omega = \langle N_\mu(E)\rangle \, \prod^g_{i=2} \alpha_i$, the total number of muons can finally be written as
\begin{equation}
  N_\mu~=~\alpha_1 \cdot \omega \ .
  \label{eq:alpha-omega}
\end{equation}

\section{\label{sec:mc-test} EAS Monte Carlo test}

\begin{figure}[t]
  \centering
  \includegraphics[width=\figsizeL\columnwidth]{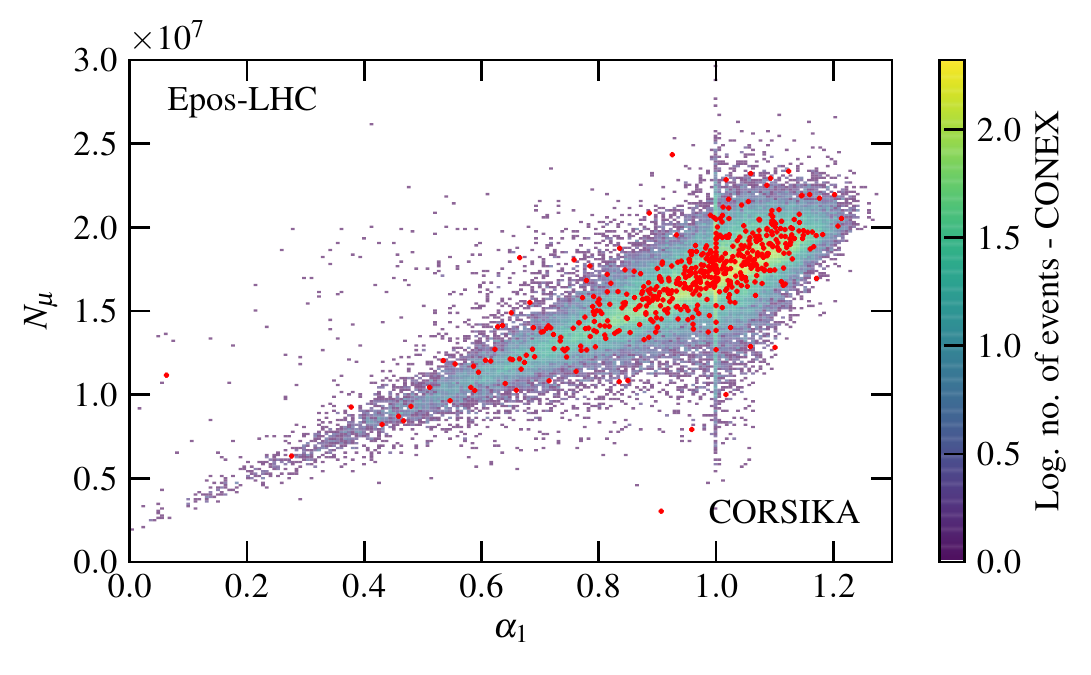}
  \caption{\label{fig:alpha-nmu} Distribution of the modified hadronic energy fraction of the first interaction, $\alpha_1$, and the number of muons, \nmu, for \conex (heat-map) and \corsika~(red points) simulations. Primaries are protons with an energy of $10^{19}\,$eV and a zenith angle of $67^{\circ}$. Hadronic interactions were simulated with \epos and \fluka. Note: the accumulation of events along the line where $\alpha_1 = 1$ corresponds to quasi-elastic/diffractive events. The corresponding distributions for \sibyll{} and \qgs{} are shown in~\ref{app:models}.
  }
\end{figure}
\begin{figure}[t]
  \centering
    \includegraphics[width=\figsize\columnwidth]{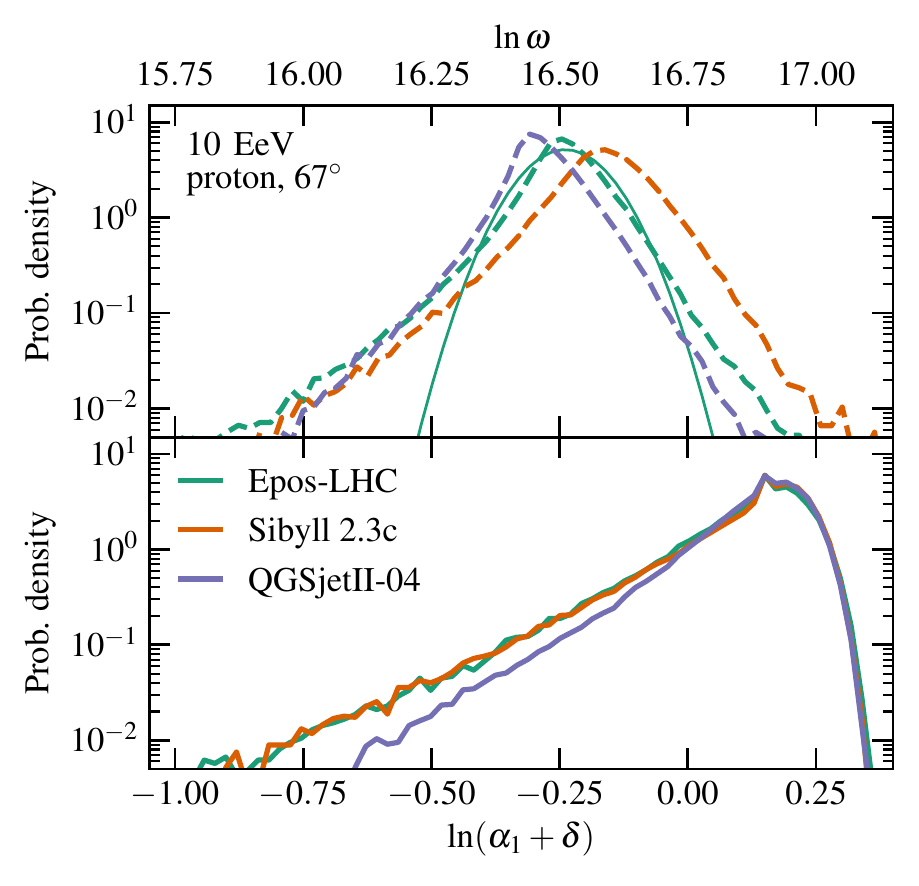}

  \caption{\label{fig:omega-alpha-dist} Average $w$-distribution ($w=\ln \omega$), showing the shower fluctuations except the ones arising from the first interaction (top,dashed lines), and $a$-distribution ($a=\ln (\alpha_1+\delta)$) of the first interaction (bottom), for different interaction models. The {\it Gaussian approach} for the $w$-distribution of \epos is also shown in a continuous line in the top panel.}
\end{figure}

\begin{figure}[t]
    \centering
  \includegraphics[width=\figsize\columnwidth]{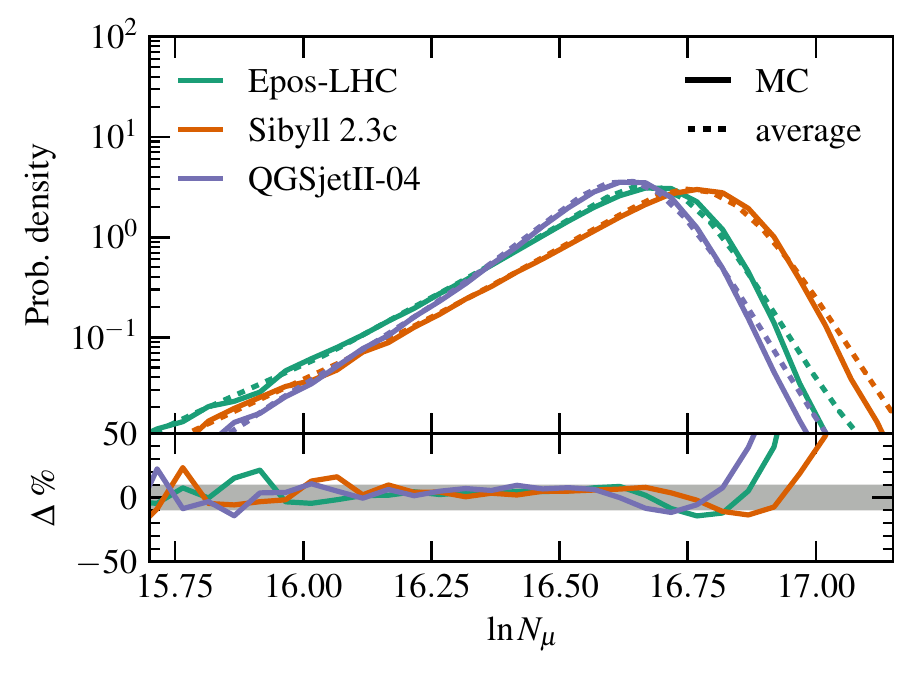}
  \caption{Distribution of the number of muons for different interaction models in MC simulations and in the model using the average approach for the $w$-distribution. $\Delta \%$ in the lower plot refers to the difference between the distributions obtained from MC and the model in percentiles of the MC result. Showers are protons at $10^{19}\,$eV with a zenith angle of $67^{\circ}$. Simulations were done with \conex. \label{fig:nmu-models} }
\end{figure}

\begin{figure}[t]
    \centering
  \includegraphics[width=\figsize\columnwidth]{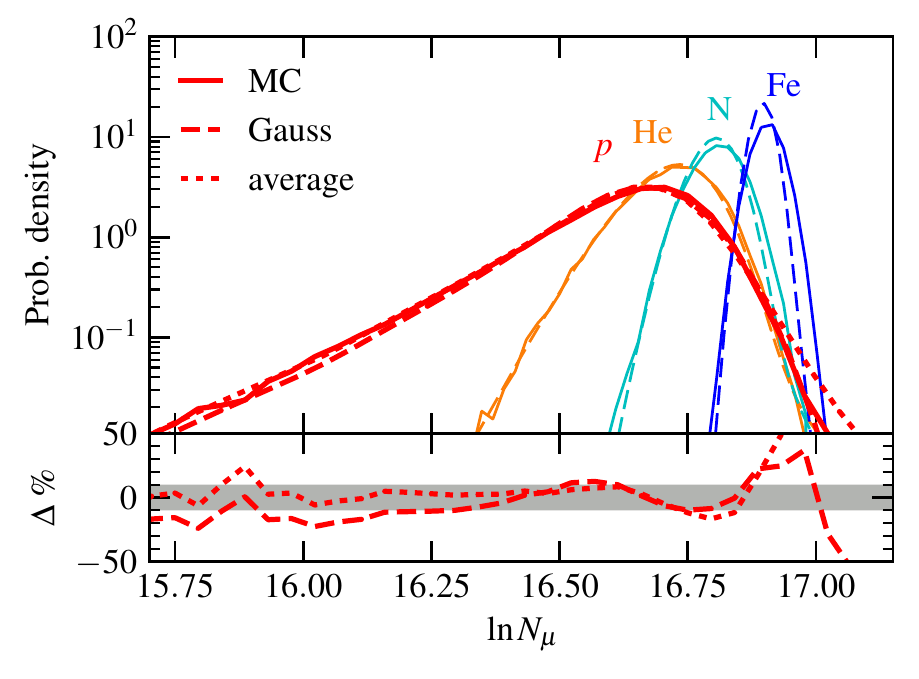}
  \caption{\label{fig:nmu-comparison} Distributions of the number of muons for different primaries from  MC, and superposition model with {\it Gaussian} approach for the $w$-distribution. For proton primaries the {\it average} approximation is also shown. The difference between the muon distribution for protons between the MC and the model in the two approximations in units of the MC is shown in the bottom of the figure ($\Delta \%$). The primary energy for the showers was $10^{19}\,$eV and the zenith angle was $67^{\circ}$. Simulations were done with \conex using \epos and \fluka. For \sibyll{} and \qgs{} see \ref{app:models}.}
\end{figure}

In order to verify our hypothesis, we have run $10^3$ full 3D-simulations of proton showers with \corsika~(v7.6400)~\cite{corsika} at a primary energy of $10^{19}\,$eV and zenith angle of $67^{\circ}$. The energy threshold for the thinning algorithm was set to $\epsilon_{\rm thin}=10^{-6}$ and weights were limited to $w_{\rm max}=  \epsilon_{\rm thin} \cdot ( E_0/1\,\mathrm{GeV})$. We have also run $10^5$ 1D-simulations of air showers with \conex~(v5.40)~\cite{Bergmann:2006yz,Pierog:2004re} for different primaries, energies $10^{16}$,$10^{17}\,$,$10^{18}\,$ and $10^{19}\,$eV and zenith angles of $38^\circ$ and $67^{\circ}$. High-energy interactions were simulated with the post-LHC models \epos~\cite{Pierog:2013ria}, \qgs~\cite{Ostapchenko:2010vb} and \sibyll{}~\cite{Riehn:2017mfm,Riehn:2015oba}, low-energies with \fluka~v2011.2c~\cite{fluka2014}. The information about the first interactions was recorded~\cite{Ulrich:2010rg}. Hadronic particles, i.e.\ particles that form the hadronic cascade and that determine $\alpha_1$, were defined as all hadrons other than $\pi^0$ and $\eta$. We refer to the number of muons at ground by $\nmu$. 

In Fig.~\ref{fig:alpha-nmu}, the  $N_\mu$ and $\alpha_1$ joint distribution, $f_{\alpha_1, \nmu}(\alpha_1, \nmu)$ derived with \conex is shown with events from \corsika superimposed. The correlation coefficient is approximately $0.8$~\footnote{$\rho_{X,Y}=cov(X,Y)/(\sigma_X \sigma_Y)$}.
For comparison, the correlation with the fraction of hadronic energy and with the multiplicity of the first interaction is about $0.6$ and $0.2$ respectively. A detailed comparison of the correlations between variables of the first interaction and the number of muons is given in~\ref{app:variables}. The correlation between $\nmu$ and the fraction of hadronic energy is discussed in~\ref{app:efrac}. Looking at the joint distribution shown in Fig.~\ref{fig:alpha-nmu} one can see that the bulk of events follows a linear relation of the form:
\begin{equation}
   \nmu~=~(\alpha_1 + \delta ) \cdot \omega \ .
  \label{eq:nmu-stat}
\end{equation}
While approximately $1/3$ of pions that are produced in a $pp$ interaction are neutral, due to the preservation of the quantum numbers of the proton (leading particle effect) the fraction of events where most of the energy is carried by a neutral pion is much smaller. In these rare cases, $\alpha_1\to 0$, and muon production is dominated by photo-pion production in the EM cascade. The additional term $\delta$ in Eq.~\eqref{eq:nmu-stat} accounts for this (compare to Eq.~\eqref{eq:alpha-omega}). Its value was found to be roughly $0.16$ across different models (\ref{app:models}). The distribution of the number of muons produced in a photon-initiated shower is shown in Fig.~\ref{fig:nmu-dis}. The vertical structure seen around $\alpha_1=1$ in Fig.~\ref{fig:alpha-nmu} corresponds to quasi-elastic/diffractive events, where a small amount of energy is transferred from the incoming primary to the target nucleus.

For convenience, we define $n\equiv\ln \nmu$, $a\equiv \ln (\alpha_1+\delta)$ and $w\equiv \ln \omega$, and thus have $n = a+w$. The probability distribution of $f_n(n)$ is then given by $f_n(n) = \int f_a(a) \, f_{w}(n-a\,|a) \, \mathrm{d}a ,$ where $f_{w}(w \, |a)$ is the $w$-distributions at different $a$, and can be accessed in simulations (see~\ref{app:details}).
By neglecting the $a$ dependence of $f_{w}(w \, |a)$, we arrive to the approximation  $f_n(n)~\simeq~\int f_a(a) \, f_{w}(n-a) \, \mathrm{d}a \ .$
In Fig.~\ref{fig:omega-alpha-dist}, $f_a(a)$ and  $f_w(w)$ are shown, where $f_w(w)\equiv \int f_w(w \, | a) \, f(a) \, \mathrm{d} a$ was  called {\it average approach}.

Fig.~\ref{fig:nmu-models} compares the {\it average} convolution approach with the exact distribution of a proton primary for different models. It can be observed that the general features are well reproduced: a) The low-$n$ tails, a reflection of the different low-$a$ tails in the models; b) The width $\sigma(n)=0.17$, which is dominated by the width $\sigma(a)=0.14$ as compared with $\sigma(w)\simeq 0.11$. Hence, $w$ plays a sub dominant role in the $\nmu$ fluctuations (notice that $f_w(w)$ includes all  contributions to the shower-to-shower fluctuations which are not accounted for by $f_a(a)$); And finally c), the different average $n$-values are a reflection of the different average values of the $w$-distribution in Fig.~\ref{fig:omega-alpha-dist}. The high-$n$ tail nevertheless would need the full correlation convolution to achieve high precision, which is out of the scope of this paper. Fig.  ~\ref{fig:nmu-comparison}, compares  also a {\it Gaussian} convolution approach, where the average and width were extracted from a Gaussian fit to  $f_w(w)$.

For a UHECR observatory to access the $a$-distribution of proton-air interactions experimentally, the shower effects would need to be unfolded from the observed $n$-distribution. This requires the use of a $f_w(w|a)$ distribution, where the $a-w$ correlations should be accounted for if one wants to attain the maximum precision. A contribution to the systematics comes from the differences in $f_w(w|a)$ across models. The present paper has laid the basis for these future analyses by demonstrating that the $n$-distribution is shaped by a more fundamental distribution of multiparticle production in the first interaction and therefore can be experimentally probed. The physics of the $a$-distribution itself and its relation with accelerator measurements is another matter for future research.

\section{The scenario of nucleus-air interactions}

In general, the UHECR composition may be composed of different types of nuclei. The shower-to-shower fluctuations of the number of muons also reflect the fluctuations of the primary UHECR mass on a shower-to-shower basis. In Fig.~\ref{fig:nmu-comparison} a comparison between MC simulations for different primaries with the mass number $A$ (number of nucleons) and the results of the superposition model are shown for the hadronic interaction model \epos. The $n$-distributions for nuclei were built from the $A$-fold self-convolution of the nucleon $n$-distribution with energy $E/A$. The gradual change in the shape of the distribution in the transition from proton showers to heavier primaries is described well by the superposition model. Fluctuations are slightly underestimated, which is expected since superposition neglects, for example, the fluctuations of the energy between the nucleon sub-showers~\cite{Dembinski:2017kpa}. For other hadronic interaction models the results are similar (see~\ref{app:mass}).

\begin{figure}[t]
    \centering
 \includegraphics[width=\figsizeL\columnwidth]{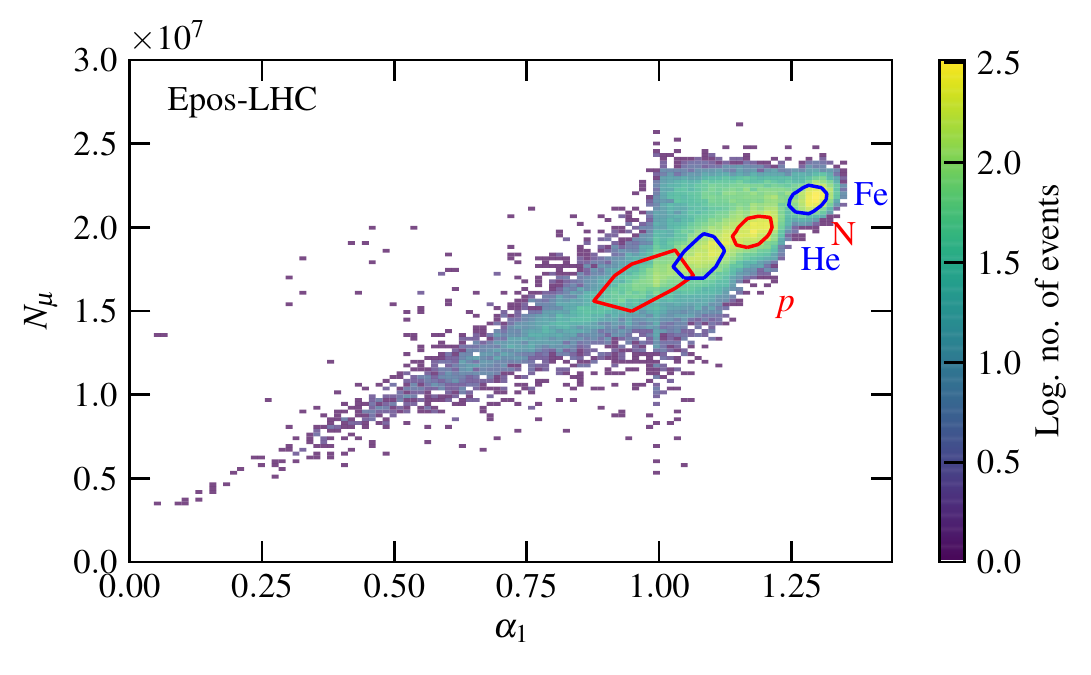}
  \caption{\label{fig:alpha-nmu-mass} Distribution of the modified hadronic energy fraction of the first interaction, $\alpha_1$ and the number of muons for different primaries. All primaries contribute equally. The contour lines enclose $1\sigma$ of the distribution of the individual primaries. The primary energy is $10^{19}\,$eV and the zenith angle is $67^{\circ}$. Simulations were done with \conex using \epos and \fluka (see~\ref{app:mass} for other models).}
\end{figure}

\begin{figure}[t]
    \centering
    \includegraphics[width=\figsize\columnwidth]{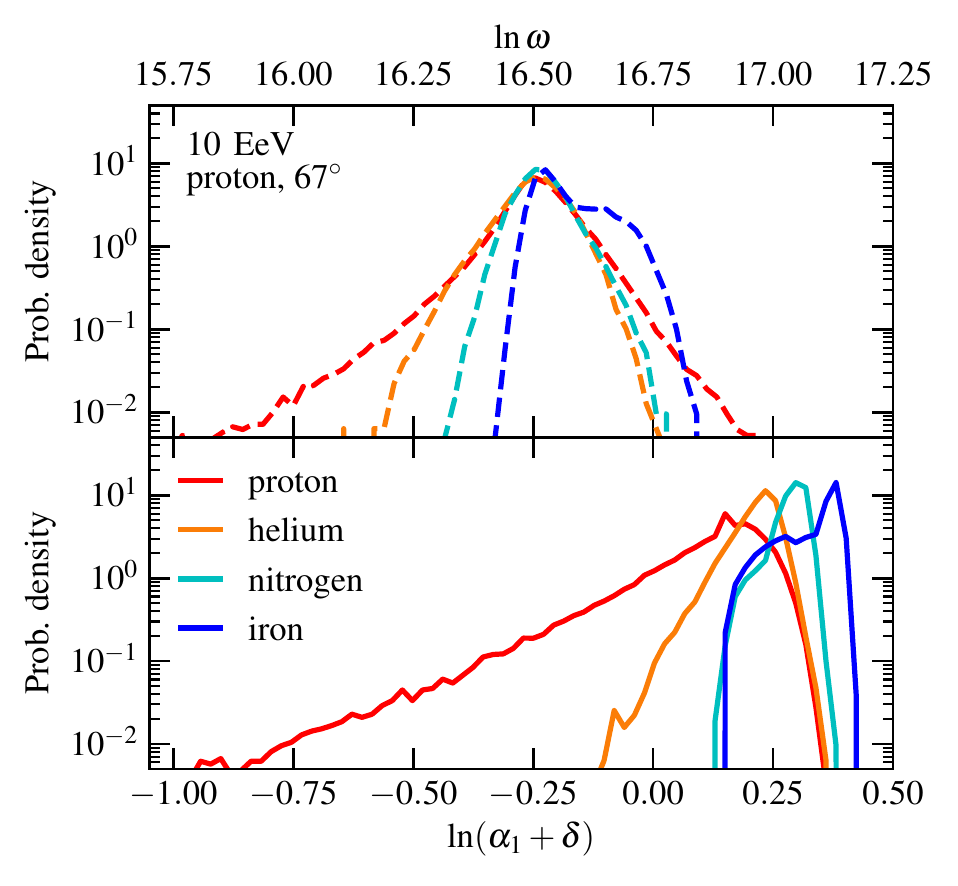}
  \caption{\label{fig:omega-alpha-dist-mass} Average $w$-distribution (top) and $a$-distribution ($a=\ln (\alpha_1+\delta)$) of the first interaction (bottom) for different primary masses. Primary energy is $10^{19}\,$eV and the zenith angle is $67^{\circ}$. Simulations were done with \conex using \epos and \fluka (see~\ref{app:mass} for other models).}
\end{figure}

Fig.~\ref{fig:alpha-nmu-mass} displays the joint-distribution of the number of muons and the modified hadronic energy fraction $\alpha_1$ after the first interaction for a mixture of different primaries. Heavier nuclei display a larger $\alpha_1$ value, reflecting the fact that, only a portion of the nucleus interacts, leaving the rest of the energy in the remaining nucleons. The number of muons increases in the same proportion, and as a consequence the populations separate. Fig.~\ref{fig:omega-alpha-dist-mass} shows the $w$-distribution and $a$-distribution. The $w$-distribution has a maximum at the same position, although the width becomes narrower as the mass number increases, while the $a$-distribution reflects the self-folding of A nucleons according to the superposition model and thus is shifted with the mass number. These figures are based on simulations with \epos. For other models the general behaviour of a shift of the populations along the diagonal in the $\nmu$-$\alpha_1$ plain with increasing mass remains the same, but details change for heavier nuclei ($A>4$) (see~\ref{app:mass}). While these details of evolution with mass may be interesting in their own right and deserve study elsewhere, they are outside the scope of this paper. The essential point here is that the tails in $n$ and $a$ are suppressed for nuclei, so that even for a mixed composition the signature tail in protons remains visible (Fig.~\ref{fig:omega-alpha-dist-mass} and Fig.~\ref{fig:nmu-comparison}). 

While there is evidence for a mixed mass composition of cosmic rays in the energy range between $10^{18.5}\,$eV and $10^{19}\,$eV~\cite{Aab:2016htd}, discussion still remains over the value of the average mass reported by different experiments~\cite{Abbasi:2014sfa,Aab:2014aea}. Nevertheless, by using a mass sensitive shower observable (like the depth at which the shower reaches its maximum developement, $\xmax$) one could experimentally select samples of events which are proton enriched, and then follow the same procedure as for the case of a pure proton UHECR beam. The unfolding of the $n$-distribution would result in a $a$-distribution that reflects the composition of the sample. If protons are present, the $a$-distribution is dominated by protons at low-$a$ values, making it possible to extract direct information of the first interaction. The same strategy was used to measure the proton-Air cross section by using $\xmax$~\cite{Auger:2012wt,Abbasi:2015fdr}.

\section{Conclusions}
In this letter, it has been demonstrated that the number of muons in EAS is connected with  a variable of the first interaction, $\alpha_1$. This variable is computed from the fraction of energy carried by the hadronic particles that sustain the hadronic cascade, being a direct probe of the hadron energy spectrum. Using this knowledge, it was shown that the shower-to-shower distribution of the muon content can be explained by the $\alpha_1$-distribution of the first interaction of the UHECR. The subsequent shower interactions give the overall scale of the muon content, but play a sub dominant role in the fluctuations.

It is worth noting that the p-Air cross section represents the only other example of a direct link between EAS measurements to a property of the first interaction of the UHECR, and which has been measured ~\cite{Auger:2012wt,Abbasi:2015fdr}. The present work demonstrates for the first time that it is possible to directly access multiparticle production variables of the p-Air interaction, which can occur at center-of-mass energies above the LHC scale.

\section*{Acknowledgments}
We want to thank the Auger-LIP group and the Auger Collaboration for inspiration and numerous discussions. In particular, to
S.\ Andringa,
H.\ Dembinski,
F.\ Diogo,
R.\ Engel,
C.\ Espirito-Santo,
T.\ Pierog,
M.\ Pimenta,
and
M.\ Unger.

We thank the financial support by  OE - Portugal, FCT, I.\ P.~, under project CERN/FIS-PAR/0023/2017.
LC and FR also thank OE - Portugal, FCT, I.\ P.~, for funding under project IF/00820/2014/CP1248/CT0001. RC is grateful for the financial support by OE - Portugal, FCT, I.\ P.~, under DL57/2016/cP1330/cT0002.

\bibliography{nmu-fluct-orig.bib}

\appendix

\section{\label{app:zenith}Shower-to-shower fluctuations in the muon content: zenith angle dependence}

\begin{figure}[!h]
    \centering
  \includegraphics[width=\figsize\columnwidth]{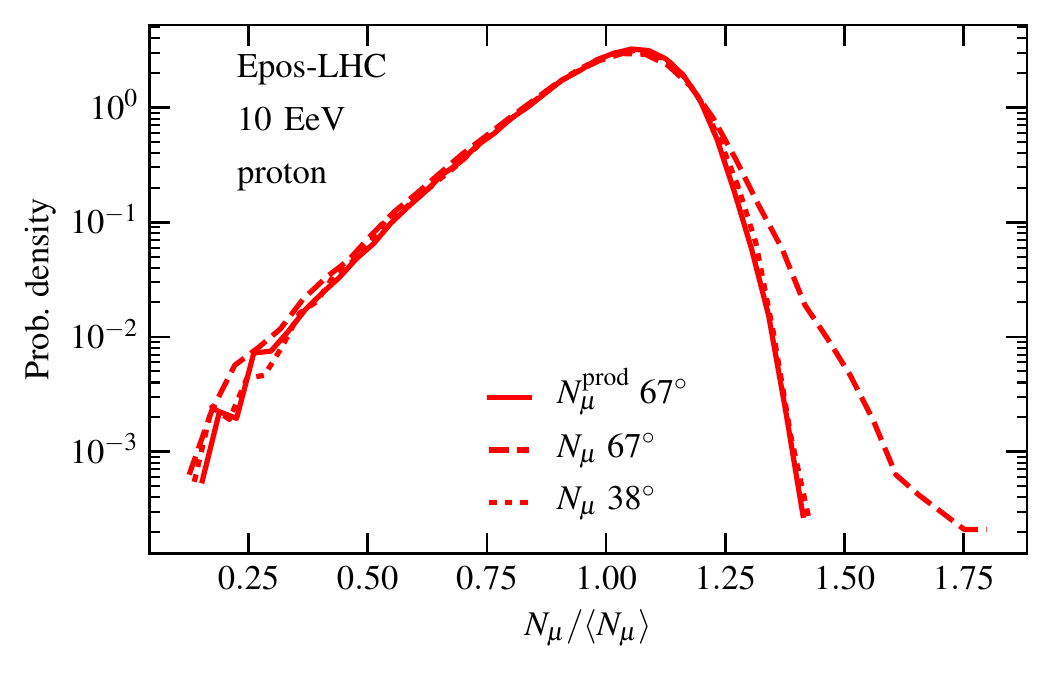}
  \caption{Distribution of the relative number of muons for proton showers at $10^{19}\,$eV and an inclination of $67\,^{\circ}$ in 1-D simulations with \conex. The number of muons at ground $N_{\mu}$ is defined all muons with $E>1\,$GeV arriving at the observation level at $1400\,$m (atmospheric depth of $2240\,$g/cm$^2$). $N_{\mu}^{\rm prod.}$ is defined as all muons with $E>1\,$GeV that were produced up to the observation level.
    \label{fig:nmu-attenuation}}
\end{figure}

The zenith angle dependence of the correlation between the number of muons at ground and at production  with $\alpha_1$ is shown in Tab.~\ref{tab:angular}. In both cases, at ground and at production, there is only a small difference in the correlation between vertical and inclined showers. The effect of the zenith angle on the distribution of the number of muons is shown in Fig.~\ref{fig:nmu-attenuation}.

\begin{table}[!h]
    \caption{\label{tab:angular} Values of the correlation for the number of muons at ground, the number of muons at production and for different zenith angles. The correlation coefficient is
      defined as $\rho_{X,Y}=cov(X,Y)/(\sigma_X \sigma_Y)$. Showers are simulated with \conex and \epos.
    }
    \begin{center}
      \renewcommand{\arraystretch}{1.5}
      \begin{tabular}{ccc}
        zenith $~\theta~$  & $~N_{\mu}~$ & $~N_{\mu}^{\rm prod.}~$ \\
        \hline 
        $38^{\circ}$ & 0.83 & 0.84 \\
        $67^{\circ}$ & 0.79 & 0.82 \\
        
      \end{tabular}
    \end{center}
\end{table}

\section{\label{app:models} Shower-to-shower fluctuations in the muon content: hadronic interaction models}

\begin{table}
  \caption{\label{tab:models} {Values of muon production parameters for different hadronic interaction models. Except $\beta$ all parameters are for proton showers at $E=10^{19}\,$eV and zenith angle of $67^\circ$. The EM contribution to muon production $\delta$ is defined by fitting $N_{\mu}=\omega \, (\alpha_1 + \delta)$. Here $\delta$ and $\omega$ are free parameters. Interaction models are \epos{}~(\textsc{Epo}), \sibyll{}~(\textsc{Sib}) and \qgs{}~(\textsc{QGS}). Model average ($\bar{x}$) and spread ($\sigma(x)$) are shown in the last column. }}
  \begin{center}
    \renewcommand{\arraystretch}{1.5}
    \begin{tabular}{ccccc}
      & \textsc{Epo} & \textsc{Sib} & \textsc{QGS} & $\bar{x}\,$ ($\sigma(x)$)\\
      \hline 
      $ \langle \ln\nmu \rangle$ & 16.63 & 16.63 & 16.60 & 16.62 (0.01) \\
      $\sigma(\nmu)/\mnmu$ & 0.155 & 0.166 & 0.138 & 0.153 (0.012) \\
      \\
      $\langle \alpha_1 \rangle$ & 0.950 & 0.953 & 0.969 & 0.957 (0.008) \\
      $\sigma(\alpha_1)$ & 0.146 & 0.152 & 0.122 & 0.14 (0.01) \\
      \\
      $\beta$ &  0.927 & 0.928 & 0.925 & 0.931 (0.002) \\
      $\delta$ & 0.163 & 0.157 & 0.161 & 0.160 (0.003) \\
    \end{tabular}
  \end{center}
\end{table}

\begin{figure*}
    \centering
  \includegraphics[width=\figsizeL\columnwidth]{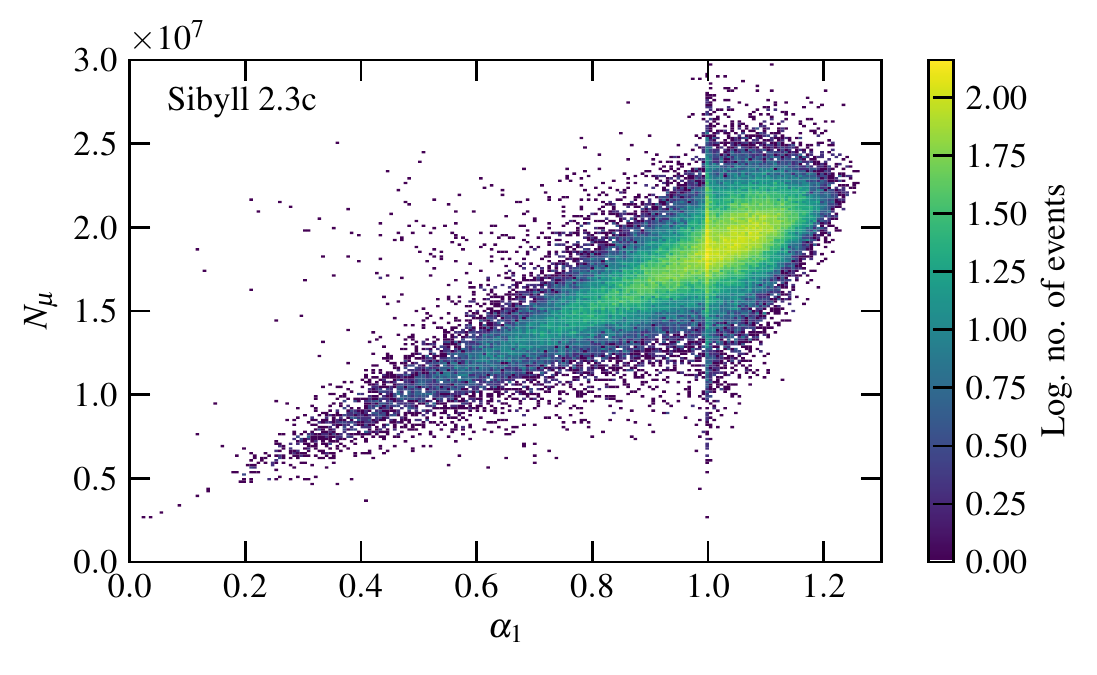}
  \hfill
  \includegraphics[width=\figsizeL\columnwidth]{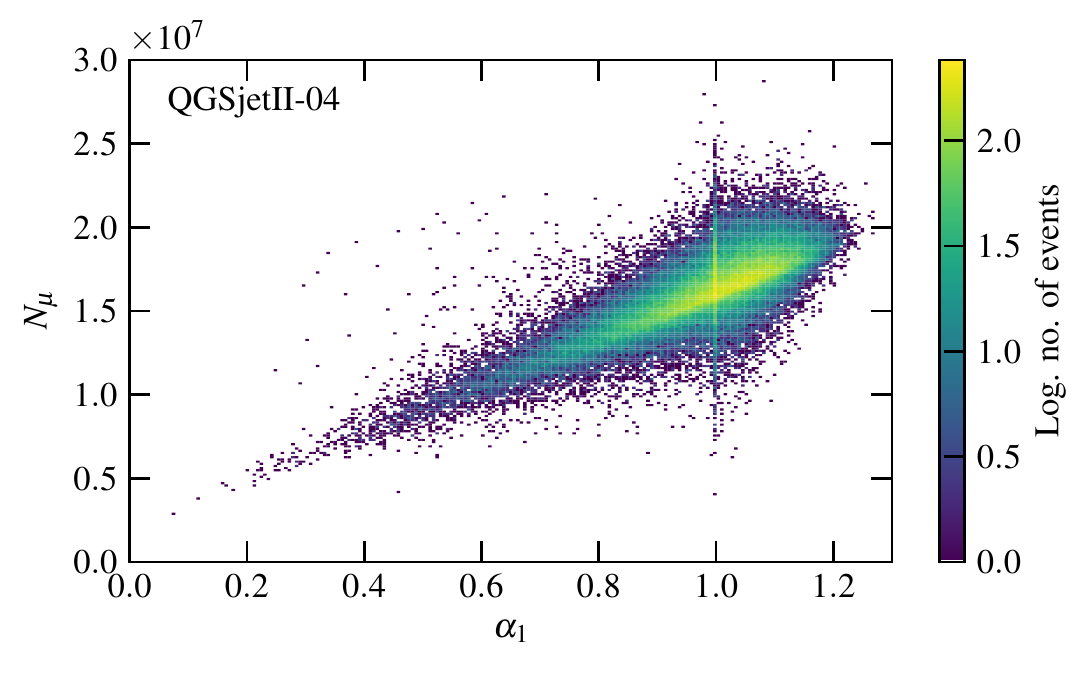}
  \caption{\label{fig:alpha-nmu-other} Correlation between $\alpha_1$ and $\nmu$ for \sibyll{} (left) and \qgs{} (right). Primaries are protons with energy $10^{19}\,$eV and zenith angle $67^{\circ}$. Showers were simulated with \conex and \fluka.}
\end{figure*}

In general we find that the results for proton showers are very similar between the interaction models. In Tab.~\ref{tab:models} the first two moments of $\alpha_1$ and $\nmu$ are compared between air shower simulations for proton primaries with the different post-LHC interaction models. The slope of the average number of muons $\beta$ and the contribution to muon production from the EM cascade, $\delta$, are shown as well. In Fig.~\ref{fig:alpha-nmu-other} the correlation between $\nmu$ and $\alpha_1$ in proton showers is shown for the interaction models \sibyll{} and \qgs{} (\epos{} is shown in Fig.~\ref{fig:alpha-nmu}).

\section{\label{app:variables} Shower-to-shower fluctuations in the muon content: variables of hadronic interactions}
Defining the correlation coefficient $\rho_{X,Y}=cov(X,Y)/(\sigma_X \sigma_Y)$, Tab.~\ref{tab:correl} shows the correlation of different variables of the first interactions with the number of muons at ground and at production. The statistical uncertainty of the correlation coefficient is given by $\sqrt{(1-\rho^2)/(N-2)}$ and is in the range of $0.02-0.04$ accross variables and interaction models. The sample size was $10^3$ and the simulations were done with \corsika.

The variables presented are: the hadronic multiplicity in the first interaction $m_1$, the total multiplicity $m_{\rm tot}$ and the fraction of energy carried by all hadrons $E_{\rm had}/E$.  A variable mixing information from multiplicity and the hadronic energy fraction is realized by $\alpha_1$. It is defined as $\alpha_1=\sum_{i=1}^{m_1}x_{i}^{\beta}$, where $\beta$ is the slope of the energy dependence of the average number of muons (see Tab.~\ref{tab:models}) and $x_{i}$ is the fraction of energy $E_i/E$ carried by hadron $i$. The fraction of energy carried by the leading hadron relative to the total energy in the hadronic cascade $E_{\rm had}$ is denoted by $\epsilon^{*}$. The inelasticity, i.e.\ the fraction of the primary energy that is carried by all other hadrons except the leading one is $\kappa_{\rm inel}$. The depth of the first interaction is $X_0$.
\begin{table}
  \caption{\label{tab:correl} Correlation coefficients $\rho_{X,Y}$
    for different variables of the first interactions with the number of muons at ground and at production in proton induced air showers. }
  \begin{center}
    \renewcommand{\arraystretch}{1.5}
    
    \begin{tabular}{cccc}
 &  & $N_{\mu}$  $(N_{\mu}^{\rm prod})$ &  \\
  & \epos & \sibyll & \qgs \\
\hline 
$\alpha_1$ & $0.79$  $(0.82)$ & $0.76$  $(0.78)$ & $0.75$  $(0.78)$ \\
$E_{\rm had}/E$ & $0.67$  $(0.66)$ & $0.67$  $(0.66)$ & $0.53$  $(0.52)$ \\
$m_1$ & $0.15$  $(0.21)$ & $0.17$  $(0.22)$ & $0.22$  $(0.27)$ \\
$\kappa_{\rm inel}$ & $-0.15$  $(-0.08)$ & $-0.11$  $(-0.07)$ & $-0.04$  $(0.00)$ \\
$m_1/m_{\rm tot}$ & $0.16$  $(0.18)$ & $0.12$  $(0.13)$ & $0.19$  $(0.18)$ \\
$X_0$ & $0.23$  $(0.12)$ & $0.21$  $(0.12)$ & $0.28$  $(0.19)$ \\
$\epsilon^*$ & $-0.01$  $(-0.08)$ & $-0.12$  $(-0.17)$ & $-0.09$  $(-0.14)$ \\
\end{tabular}

  \end{center}    
\end{table}

\section{\label{app:efrac} Shower-to-shower fluctuations in the muon content: comparison of $\alpha_1$ and $E_{\rm had}/E$}
The modified energy fraction $\alpha_1$ contains more information on particle production and therefore exhibits a stronger correlation with muon production than the pure energy fraction of hadrons $E_{\rm had}/E=\sum_i^{m} x_{i}$. On the other hand $E_{\rm had}/E$ is strictly constrained to the interval between 0 and 1 by energy conservation, while $\alpha_1$ is only constrained by the multiplicity. Given a maximal multiplicity $m_{\rm max}$ and assuming energy is shared equally ($x_{i}=1/m_{\rm max}$), the maximum of $\alpha_1$ is
\begin{equation}
  \alpha_{\rm max}=\sum_i^{m_{\rm max}} x^{\beta}_{i}= \sum_i^{m_{\rm max}} \left (\frac{1}{m_{\rm max}} \right)^{\, \beta}= m_{\rm max}^{1-\beta} \ .
\end{equation}
With $m_{\rm max}=1000$ and $\beta=0.93$ this gives $\alpha_{\rm max}\sim 1.6$.

\begin{figure}
    \centering
  \includegraphics[width=\figsizeL\columnwidth]{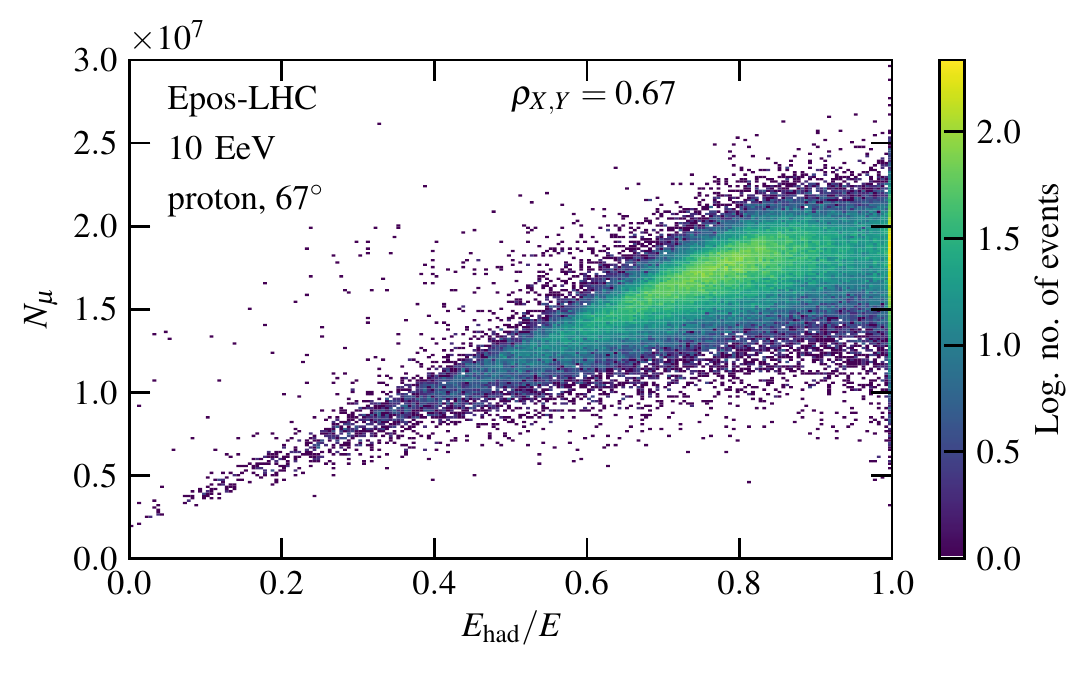}
  \caption{Distribution of the hadronic energy fraction of the first interaction, $E_{\rm had}/E$, and the number of muons, \nmu in proton showers with a primary energy of $10^{19}\,$eV and zenith angle of $67^{\circ}$. Simulations were done with \conex using \epos and \fluka.
    \label{fig:ehad-nmu}}
\end{figure}

\begin{figure}
    \centering
  \includegraphics[width=\figsizeL\columnwidth]{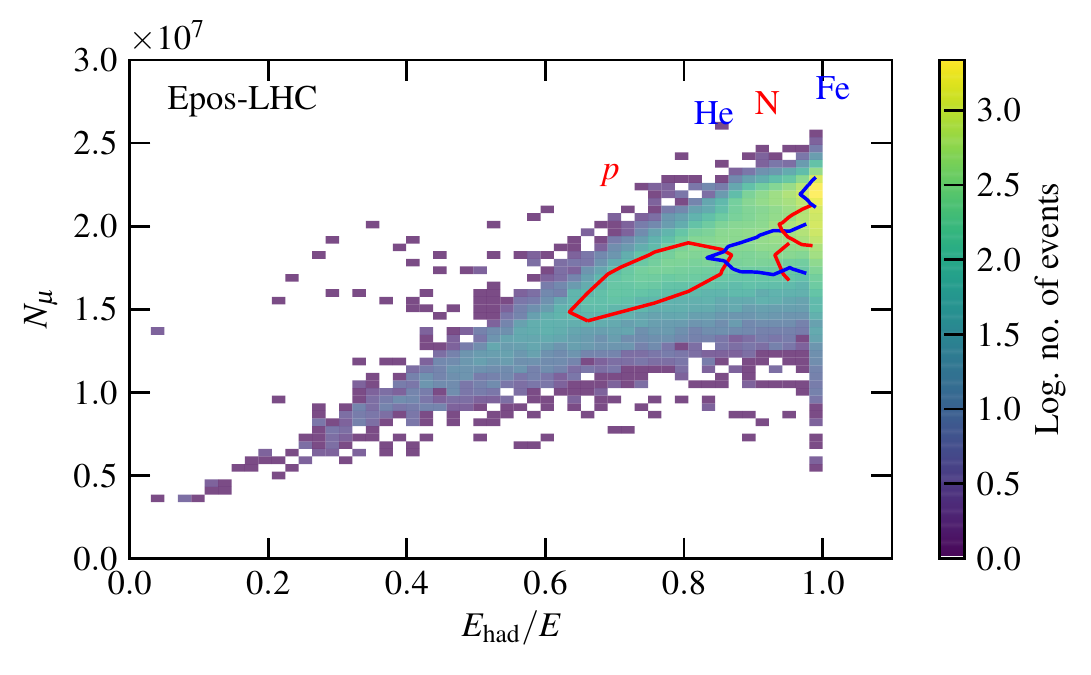}
  \caption{Distribution of the hadronic energy fraction of the first interaction, $E_{\rm had}/E$, and the number of muons, \nmu for different primaries. Primary energy is $10^{19}\,$eV and the zenith angle is $67^{\circ}$. Simulations were done with \conex using \epos and \fluka.
    \label{fig:ehad-nmu-mass}}
\end{figure}

In Fig.~\ref{fig:ehad-nmu} the correlation between the number of muons and the hadronic energy fraction is shown for proton showers at $10^{19}\,$eV and zenith angles of $67^{\circ}$. In Fig.~\ref{fig:ehad-nmu-mass} the correlation is shown for mixed composition. 

\section{\label{app:mass} The scenario of nucleus-air interactions: hadronic interaction models}
\begin{figure*}
    \centering
  \includegraphics[width=\figsizeL\columnwidth]{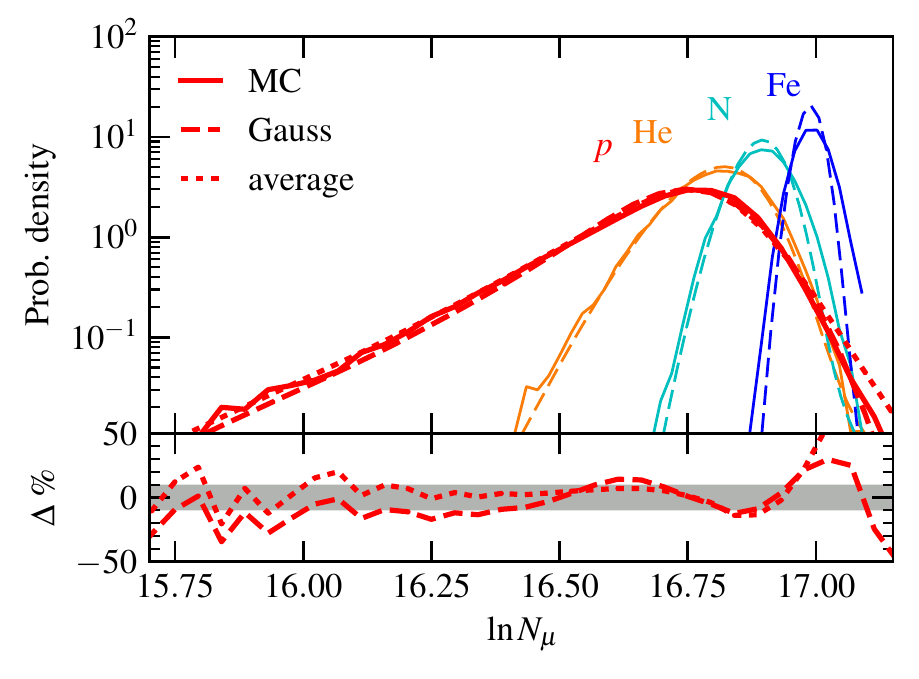}
  \hfill
  \includegraphics[width=\figsizeL\columnwidth]{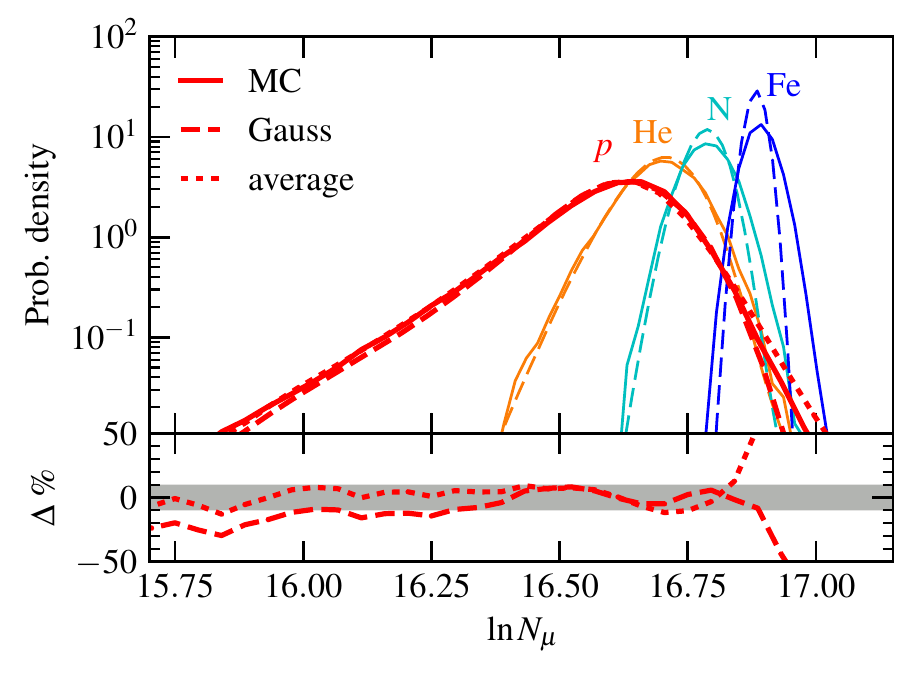}
  \caption{\label{fig:nmu-distr-super} Comparison of the distributions of the number of muons from MC and superposition model with {\it Gaussian} approach for the $w$-distribution for \sibyll{}~(left) and \qgs{} (right) (\epos is shown in Fig.~\ref{fig:nmu-comparison}). For proton primaries the {\it average} approximation is also shown. The difference between the muon distribution for protons between the MC and the model in the two approximations in units of the MC is shown in the bottom of the figure ($\Delta \%$). Primary energy is $10^{19}\,$eV and zenith angle $67^{\circ}$. Showers were simulated with \conex and \fluka.}
\end{figure*}

\begin{figure*}
    \centering
  \includegraphics[width=\figsize\columnwidth]{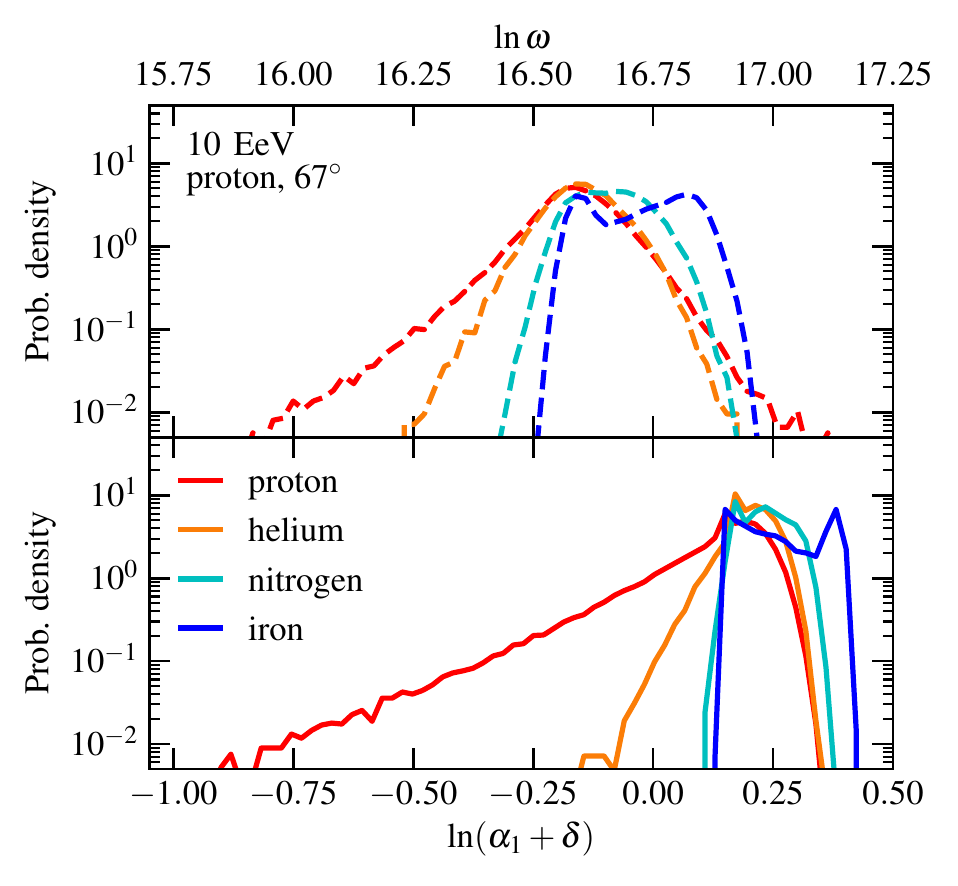}
  \hfill
  \includegraphics[width=\figsize\columnwidth]{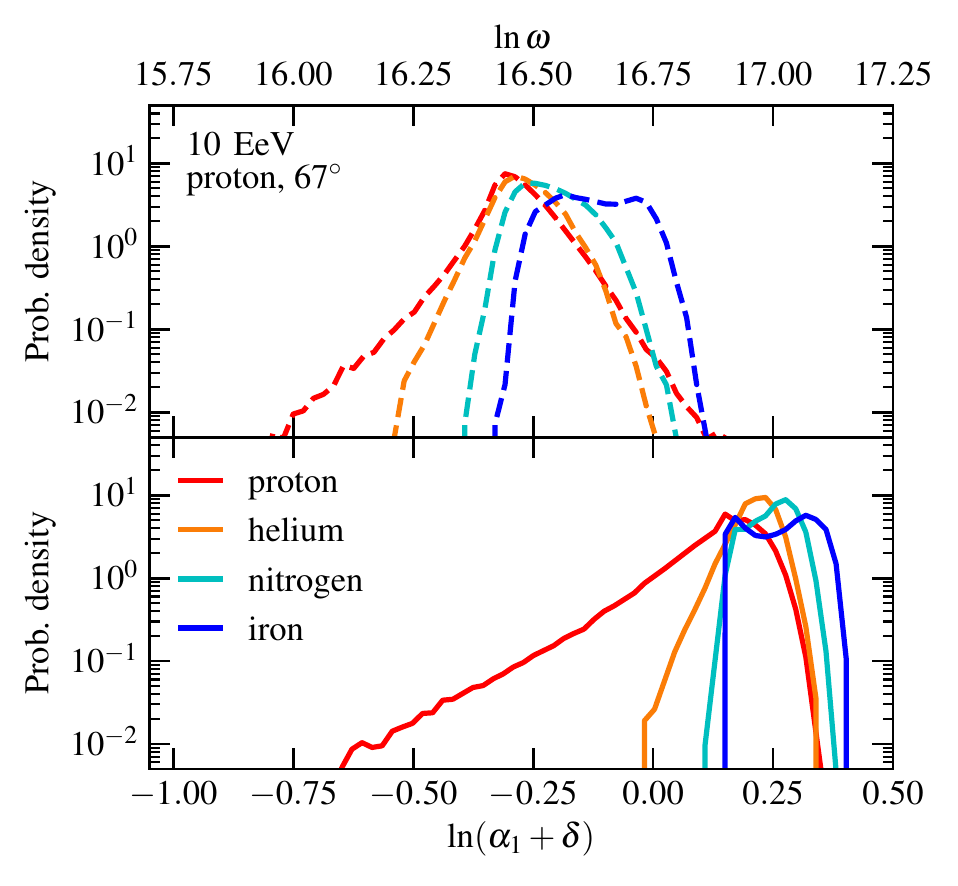}
  \caption{\label{fig:omega-alpha-dist-mass2} Average $w$-distribution (top) and $a$-distribution ($a=\ln (\alpha_1+\delta)$) of the first interaction (bottom) for different primary masses (corresponding to Fig.~\ref{fig:omega-alpha-dist-mass}). Primary energy is $10^{19}\,$eV and the zenith angle is $67^{\circ}$. Simulations were done with \conex using \sibyll~(left) and \qgs~(right) and \fluka.}
\end{figure*}

\begin{figure*}
    \centering
  \includegraphics[width=\figsizeL\columnwidth]{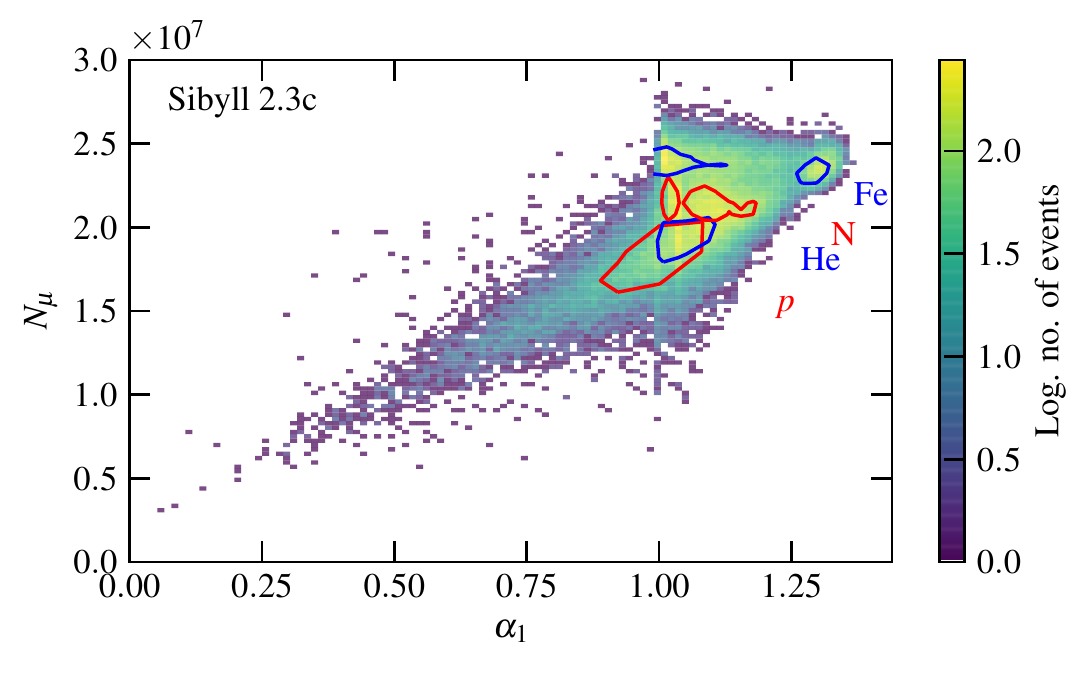}
  \hfill
  \includegraphics[width=\figsizeL\columnwidth]{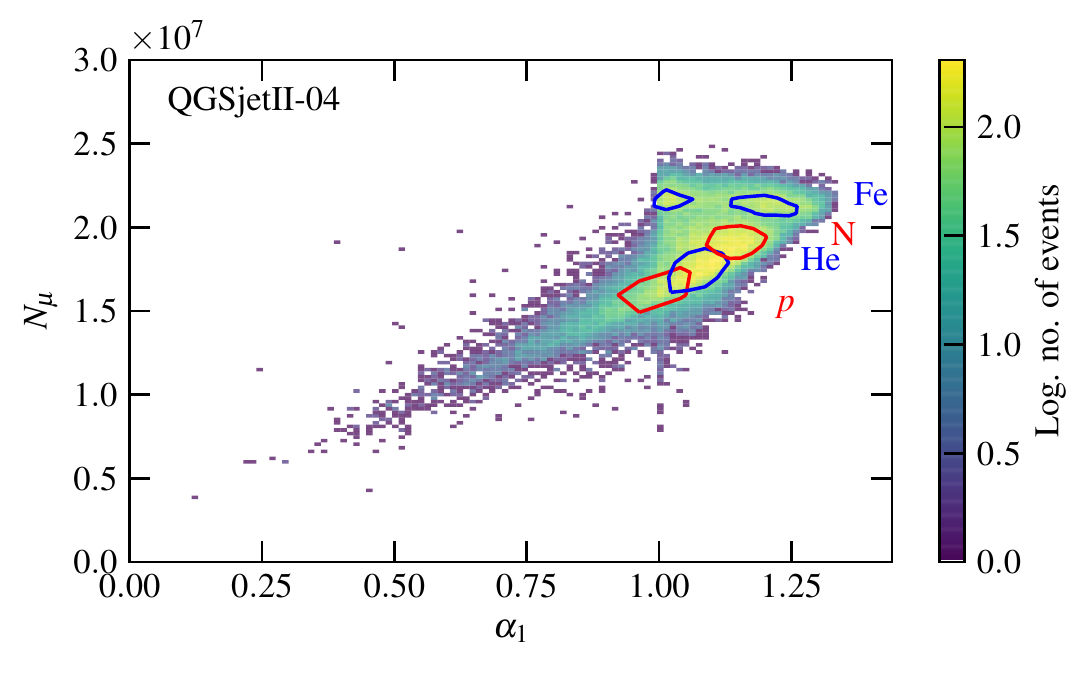}
  \caption{\label{fig:alpha-nmu-mass2} Distribution of the modified hadronic energy fraction of the first interaction, $\alpha_1$ and the number of muons for a mixed composition of primaries. Hadronic interaction models are: \sibyll~(left) and \qgs~(right) (\epos is shown in Fig.~\ref{fig:alpha-nmu-mass}). The contour lines enclose $1\sigma$ of the distribution of the individual primaries. The primary energy is $10^{19}\,$eV and the zenith angle is $67^{\circ}$. Simulations were done with \conex and \fluka.}
\end{figure*}

In Tab.~\ref{tab:models-mass} the evolution of the parameters of muon production for the hadronic interaction models \epos, \sibyll{} and \qgs{} as a function of the primary mass is shown. As the superposition model predicts the average number of muons increases with the mass while the relative fluctuations decrease. This behaviour is confirmed by Fig.~\ref{fig:nmu-distr-super} (Fig.~\ref{fig:nmu-comparison} for \epos) which shows the full distribution of $\nmu$ for air showers at $10^{19}\,$eV. For $\alpha_1$ the mean shows a similar monotonous increase. The fluctuations of $\alpha_1$ on the other hand first decrease with the number of nucleons and then increase again. In Fig.~\ref{fig:omega-alpha-dist-mass2} (and Fig.~\ref{fig:omega-alpha-dist-mass} for \epos) the evolution of the distributions of $a=\ln \alpha_1$ and $w=\ln \omega$ with mass is shown. Since $\alpha_1$ depends on energy and multiplicity, it is sensitive to the fragmentation of the nucleus in the atmosphere. The increasingly larger difference between the shapes of the $\alpha_1$-distributions for the different interaction models in Fig.~\ref{fig:omega-alpha-dist-mass2} is probably due to the different approaches used to treat nuclear fragmentation in the models. The joint-distribution of $\alpha_1$ and $\nmu$ is shown in Fig.~\ref{fig:alpha-nmu-mass2} (\epos in Fig.~\ref{fig:alpha-nmu-mass}).

\begin{table}
  \caption{\label{tab:models-mass} Values of muon production parameters for different primaries at $10^{19}\,$eV and zenith angle of $67^{\circ}$. Simulations done with \conex and \fluka.}
  \begin{center}
    \renewcommand{\arraystretch}{1.5}
    \begin{tabular}{cccccc}
      & & p & He & N & Fe \\
      \hline
                                 & \epos   & 16.61 & 16.72 & 16.81 & 16.91 \\
      $ \langle \ln\nmu \rangle$ & \sibyll & 16.70 & 16.80 & 16.90 & 16.99 \\
                                 & \qgs    & 16.59 & 16.70 & 16.79 & 16.89 \\
      \\
                                             & \epos   & 0.16 & 0.08 & 0.05 & 0.03 \\
      $ \sigma(\nmu) / \langle \nmu \rangle$ & \sibyll & 0.16 & 0.09 & 0.05 & 0.03 \\
                                             & \qgs    & 0.13 & 0.07 & 0.05 & 0.03 \\

      \\
                                  & \epos   & 0.96 & 1.08 & 1.17 & 1.23 \\
      $ \langle \alpha_1 \rangle$ & \sibyll & 0.97 & 1.07 & 1.11 & 1.15 \\
                                  & \qgs    & 0.98 & 1.08 & 1.13 & 1.14 \\
      \\
                                  & \epos   & 0.14 & 0.06 & 0.04 & 0.09 \\
      $ \sigma(\alpha_1)$         & \sibyll & 0.14 & 0.06 & 0.06 & 0.11 \\
                                  & \qgs    & 0.12 & 0.06 & 0.06 & 0.09 \\
    \end{tabular}
  \end{center}
\end{table}

\section{\label{app:details} Details of the $\alpha_1$-$\omega$ distributions}

\begin{figure}
    \centering
  \includegraphics[width=\figsize\columnwidth]{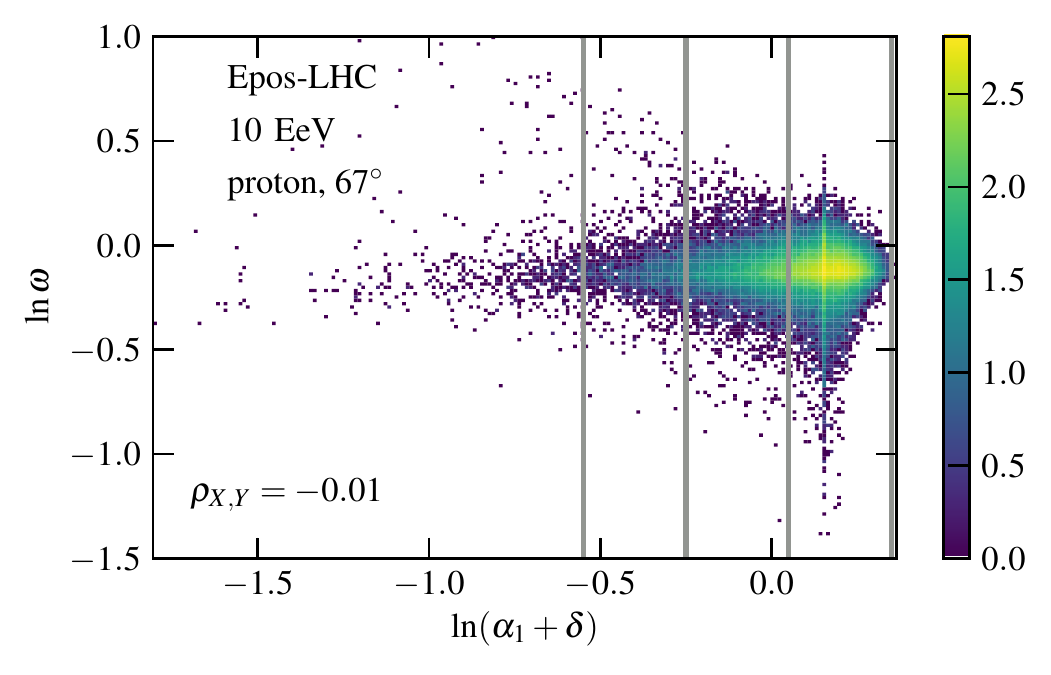}
  \caption{Two dimensional distribution of $\ln \omega$ and $\ln(\alpha + \delta)$. The overall correlation is low, however there are local structures. Projections onto $\ln \omega$ in the three regions enclosed by the grey lines are shown in the next panel. Simulations are done with \conex for proton showers at $10^{19}\,$eV and inclination of $67^{\circ}$.
    \label{fig:omega-alpha-slices}}
\end{figure}

\begin{figure*}
    \centering
  \includegraphics[width=\figsize\columnwidth]{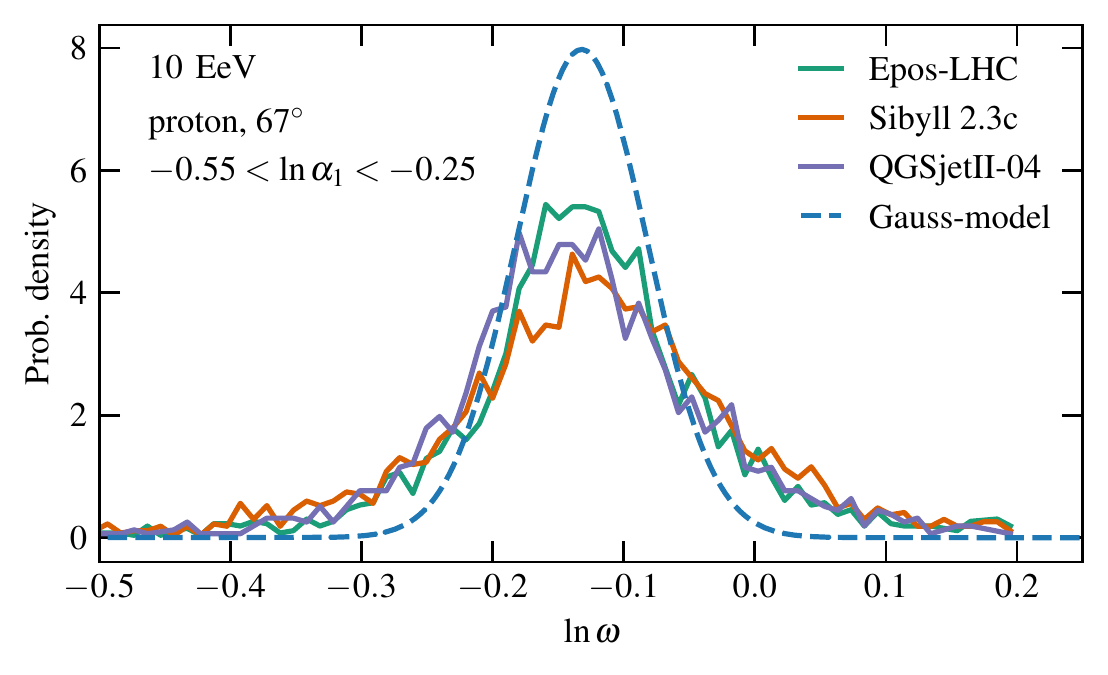}
  \hfill
  \includegraphics[width=\figsize\columnwidth]{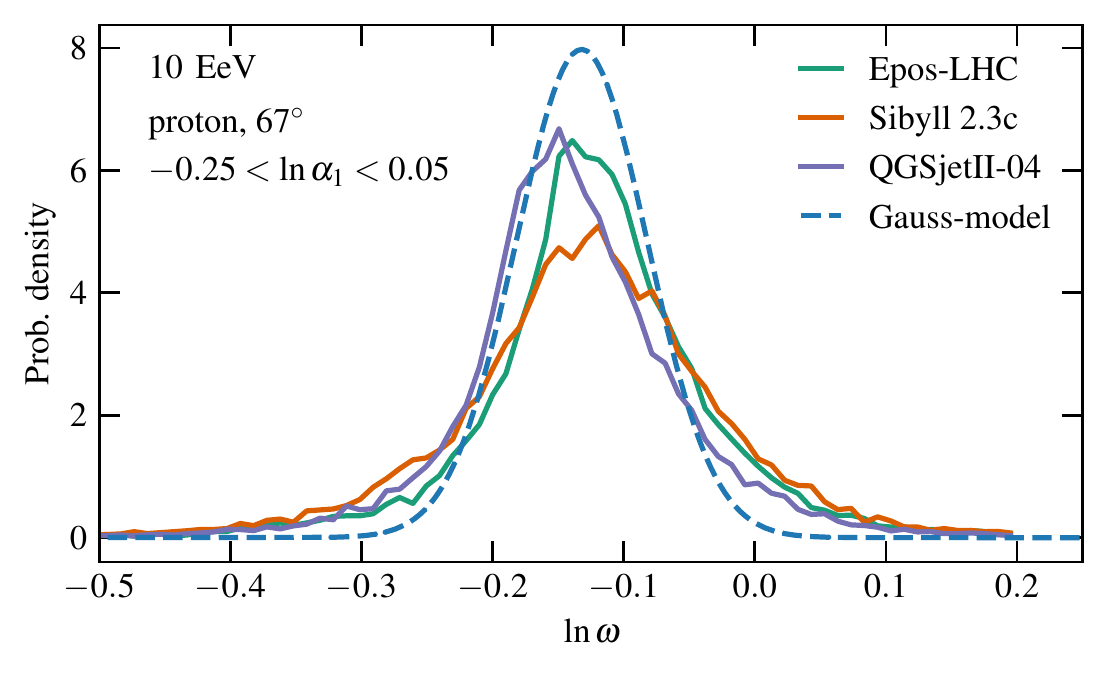}

  \caption{Distribution of $\ln \omega$ in different regions of $\ln(\alpha+\delta)$. A comparison of the interaction models for the regions with low (left) and intermediate (right) $\alpha$ is shown. For each panel the approximation of $f_{\ln \omega}(\ln \omega)$ with a Gaussian is shown as well. Showers are simulated with \conex.
    \label{fig:omega-slices}}
\end{figure*}

\begin{figure*}
    \centering
    \includegraphics[width=\figsize\columnwidth]{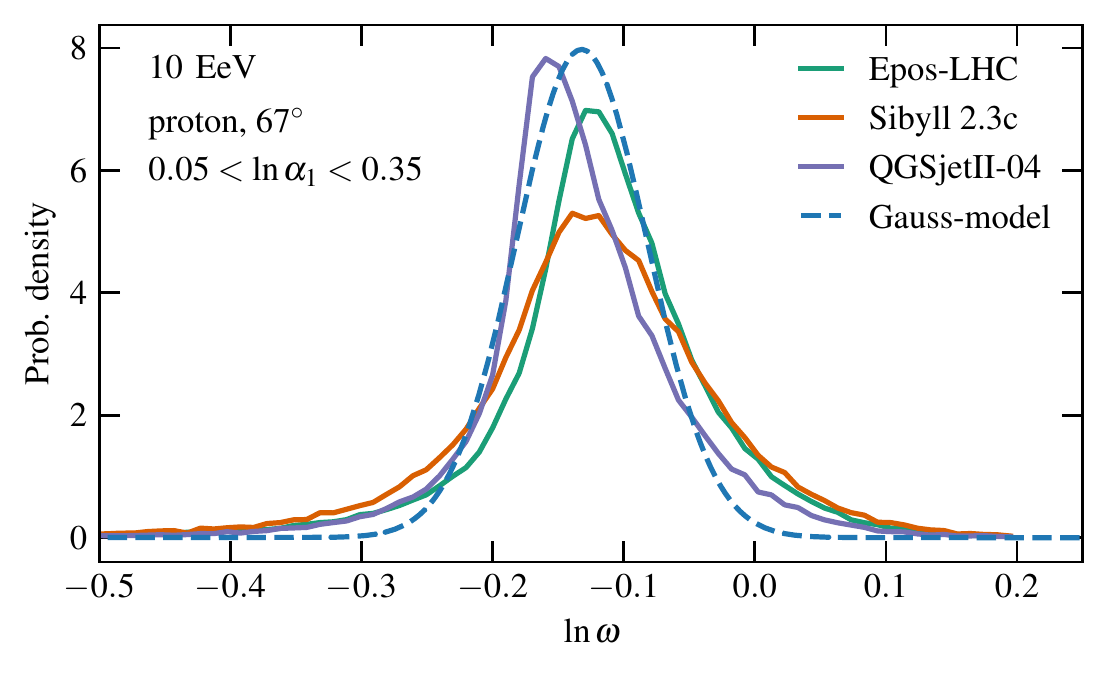}
  \hfill
  \includegraphics[width=\figsize\columnwidth]{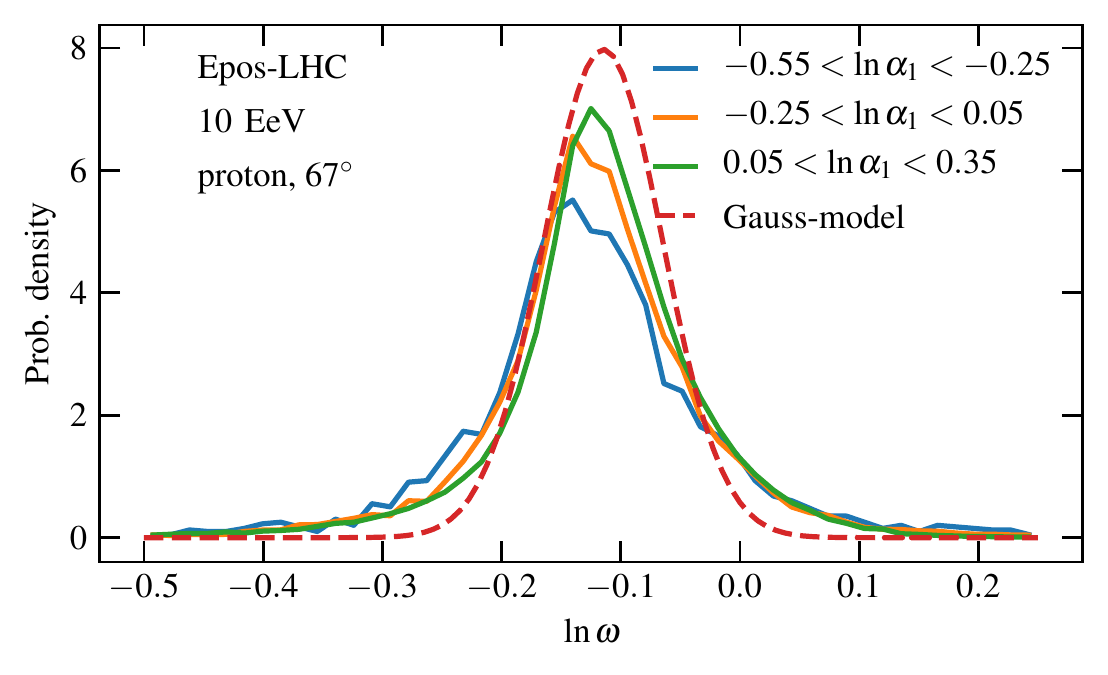}
  \caption{Distribution of $\ln \omega$ in the region of large $\alpha$ for different interaction models (left). Comparison of the distribution for \epos in the different regions is show on the right. For each panel the approximation of $f_{\ln \omega}(\ln \omega)$ with a Gaussian is shown. Showers are simulated with \conex.
    \label{fig:omega-slices2}}
\end{figure*}

In Fig.~\ref{fig:omega-alpha-slices} the distribution of $f_{w,a}(w,\, a)$ (see Sec.~\ref{sec:mc-test}) is shown for proton showers with $E=10^{19}\,$eV and zenith angle of $67^{\circ}$ that were simulated with \epos. The remaining correlation between $a$ and $w$ is $0.01$. The vertical grey lines indicate the slices along $a$ that are shown in Fig.~\ref{fig:omega-slices} and Fig.~\ref{fig:omega-slices2}.

\end{document}